\documentclass[preprint,review,12pt]{elsarticle}

\usepackage{lineno,hyperref}
\modulolinenumbers[5]

\journal{Materials Today Communications}

\usepackage{graphicx}
\usepackage{color}

\usepackage{amssymb}
\usepackage{amsmath}

\begin{document}

\begin{frontmatter}

\title{The influence of Zn$^{2+}$ ions on the local structure and thermochromic properties of Cu$_{1-x}$Zn$_x$MoO$_4$ solid solutions}

\author[ISSP]{Inga Pudza\corref{ip}}
\cortext[ip]{Corresponding author}
\ead{inga.pudza@cfi.lu.lv}

\author[ISSP]{Andris Anspoks}

\author[ISSP]{Arturs Cintins}

\author[ISSP,DESY]{Aleksandr Kalinko}

\author[DESY]{Edmund Welter}

\author[ISSP]{Alexei Kuzmin\corref{ak}}
\cortext[ak]{Corresponding author}
\ead{a.kuzmin@cfi.lu.lv}

\address[ISSP]{Institute of Solid State Physics, University of Latvia, Kengaraga Street 8, LV-1063 Riga, Latvia}
\address[DESY]{Deutsches Elektronen-Synchrotron (DESY) -- A Research Centre of the Helmholtz Association, Notkestrasse 85, D-22607 Hamburg, Germany}

\begin{abstract}
The influence of zinc ions on the thermochromic properties of polycrystalline Cu$_{1-x}$Zn$_x$MoO$_4$ ($x$=0.10, 0.50, 0.90) solid solutions was studied by X-ray absorption spectroscopy at the Cu, Zn and Mo K-edges. 
Detailed structural information on the local environment of metal ions was obtained from 
the simultaneous analysis of EXAFS spectra measured at three metal absorption edges using the reverse Monte Carlo method.  Thermochromic phase transition  with the hysteretic behaviour between $\alpha$ and $\gamma$ phases was observed in Cu$_{0.90}$Zn$_{0.10}$MoO$_4$ solid solution.
It was found that the local environment of molybdenum ions is most susceptible to the substitution of copper for zinc and, upon cooling, transforms from tetrahedral MoO$_4$ to distorted octahedral MoO$_6$.  

\end{abstract}


\end{frontmatter}


\newpage

\section{Introduction}
\label{intro}

Thermochromic materials change their colour in dependence on temperature but the nature of this effect may have a different origin \cite{Day1968,Kiri2010,Wang2016}. In solid inorganic materials, a gradual colour change may occur due to the band gap variation induced by lattice expansion/contraction, while a distinct colour change may take place upon a phase transition affecting the material structure.
Both types of thermochromic behaviour are found in copper molybdate (CuMoO$_4$) 
\cite{Steiner2001,Gaudon2007a,Jonane2019b,Joseph2020}. 
Moreover, it has been demonstrated that replacing molybdenum with tungsten or copper with other divalent metals makes it possible to change the thermochromic response \cite{Gaudon2007a,Gaudon2007b,Gaudon2010,Yanase2013,Blanco2015,Robertson2015,Yanase2019}.  Thermochromic properties have recently also been found in other molybdates such as NiMoO$_4$ \cite{DaSilva2020} and CoMoO$_4$ \cite{Costa2020}. 

At ambient pressure, pure CuMoO$_4$ exists in two triclinic (space group $P\bar{1}$) phases: high-temperature  green $\alpha$-CuMoO$_4$ and low-temperature brownish $\gamma$-CuMoO$_4$ \cite{Wiesmann1997}. The $\gamma$-CuMoO$_4$ (stable also at high pressure)  is composed of distorted CuO$_6$ and MoO$_6$ octahedra. During the structural phase transition to $\alpha$-phase, $1/3$ of the octahedra CuO$_6$ transforms to the square-pyramidal CuO$_5$, and all molybdenum atoms become tetrahedrally coordinated by oxygen atoms. The thermochromic phase transition between $\alpha$ and $\gamma$-phases is of the first order and demonstrates a hysteretic behaviour \cite{Gaudon2007a,Wiesmann1997,Jonane2018b}. 

$\alpha$-ZnMoO$_4$  has a white color and is isostructural at ambient conditions  to $\alpha$-CuMoO$_4$ \cite{Abrahams1967,Yadav2019}. It is composed of distorted ZnO$_6$ octahedra, ZnO$_5$ square-pyramids and MoO$_4$ tetrahedra \cite{Abrahams1967}. The $\alpha$-phase of ZnMoO$_4$ is stable in the whole temperature range till it decomposes at 1280~K \cite{Abrahams1967, Steiner2003}.
Note that there also exists  $\beta$-ZnMoO$_4$ phase, which adopts the wolframite structure type with the monoclinic space group $P2/c$, where both Zn and Mo atoms are octahedrally bonded to six oxygen atoms \cite{Cavalcante2013,MtiouiSghaier2015}. 

The proximity of the Cu$^{2+}$ (0.73~\AA) and Zn$^{2+}$ (0.74~\AA) ionic radii favors a substitution, so that  a complete solid solution series Cu$_{1-x}$Zn$_x$MoO$_4$  (0$\leq$x$\leq$1) exists \cite{Steiner2003,Reichelt2000}. 
Diffraction data reported in \cite{Steiner2003} show that Zn content higher than $\sim$17.5\% leads to a phase similar to $\alpha$-ZnMoO$_4$.

The substitution of copper with zinc produces a strong effect on thermochromic properties of copper molybdate by tuning its band gap from 2.78~eV in $\alpha$-CuMoO$_4$ to 3.6~eV in $\alpha$-ZnMoO$_4$ \cite{Tiwari2020}.  

Magnetic susceptibility measurements of Cu$_{1-x}$Zn$_x$MoO$_4$ (0$\leq$x$\leq$0.1) reported in \cite{Asano2011} indicate a shift of $\alpha$ $\leftrightarrow$ $\gamma$  phase transition hysteresis to the lower temperatures by about 75~K with increasing Zn$^{2+}$ content. 

Yanase \textit{et al.} \cite{Yanase2019} studied the thermochromic phase transition in the Cu$_{1-x}$Zn$_x$Mo$_{0.94}$W$_{0.06}$O$_4$ solid solutions with an even smaller amount of Zn (0$\leq$x$\leq$0.05). The results show that the substitution of Cu$^{2+}$ with Zn$^{2+}$ reduces the phase transition temperature of CuMo$_{0.94}$W$_{0.06}$O$_4$ \cite{Yanase2013}. 
For instance, the Zn content of $x$=0.01 induces the largest colour change in the temperature range of 303--343~K in the visible region due to an increase of the $\gamma$-to-$\alpha$ phase transition temperature \cite{Yanase2019}. 

A deeper understanding of the relationship between crystal structure and thermochromic properties in Cu$_{1-x}$Zn$_x$MoO$_4$ solid solutions requires detailed knowledge of the short-range order variations upon transition metal ion substitution and temperature change.  X-ray absorption spectroscopy (XAS) is a suitable tool to address this problem because it provides information about the local environment around atoms of a specific type. However, the large number of parameters required to describe such complex and low-symmetric molybdate structures makes the conventional extended X-ray absorption fine structure (EXAFS) analysis ineffective. 
At the same time, the EXAFS analysis based on the reverse Monte-Carlo (RMC) method coupled with an evolutionary algorithm (EA) approach \cite{Timoshenko2014} is an effective method that provides the possibility of obtaining a reliable single structural model by fitting EXAFS spectra from several absorption edges simultaneously.

Indeed, the RMC/EA method has been used by us in the past to demonstrate the sensitivity of XAS to the local structure variations in CuMo$_{1-x}$W$_x$O$_4$ upon the thermochromic phase transition  \cite{Jonane2020}. Moreover, the hysteretic behaviour of the transition has been evidenced by the analysis of the Mo K-edge  X-ray absorption near-edge structure (XANES) and EXAFS \cite{Jonane2018b,Jonane2018c,Jonane2020} and the W L$_3$-edge high-energy resolution fluorescence detected XANES (HERFD-XANES) \cite{Pudza2021}. 

In this study, we applied the temperature-dependent  XANES/EXAFS spectroscopy at the Cu, Zn and Mo K-edges to investigate the influence of zinc ions on the local structure and lattice dynamics of ZnMoO$_4$ and Cu$_{1-x}$Zn$_{x}$MoO$_4$ ($x$=0.10, 0.50, 0.90)  solid solutions. 
The hysteretic thermochromic phase transition between $\alpha$ and $\gamma$ phases was found only in Cu$_{0.90}$Zn$_{0.10}$MoO$_4$ and is explained by the instability of molybdenum coordination under lattice volume change induced by the substitution of copper with zinc and temperature.

\section{Experimental details and data analysis}
\label{exper}

Polycrystalline Cu$_{1-x}$Zn$_x$MoO$_4$ ($x$=0.10, 0.50, 0.90) solid solutions and pure CuMoO$_4$ were synthesized using solid-state reaction method by heating a mixture of CuO, ZnO and MoO$_3$ powders in a stoichiometric amount at 650$^\circ$C in air for 8 hours, followed by natural cooling to room temperature. Pure $\alpha$-ZnMoO$_4$ was synthesized also by solid-state method but a mixture of ZnO and MoO$_3$ powders was heated for 15 hours. The colour of the resulting samples changes from white ($x$=1) to greenish ($x$=0.1) with intermediate pale green colours.  

\begin{figure}[t]
	\centering
	\includegraphics[width=0.8\textwidth]{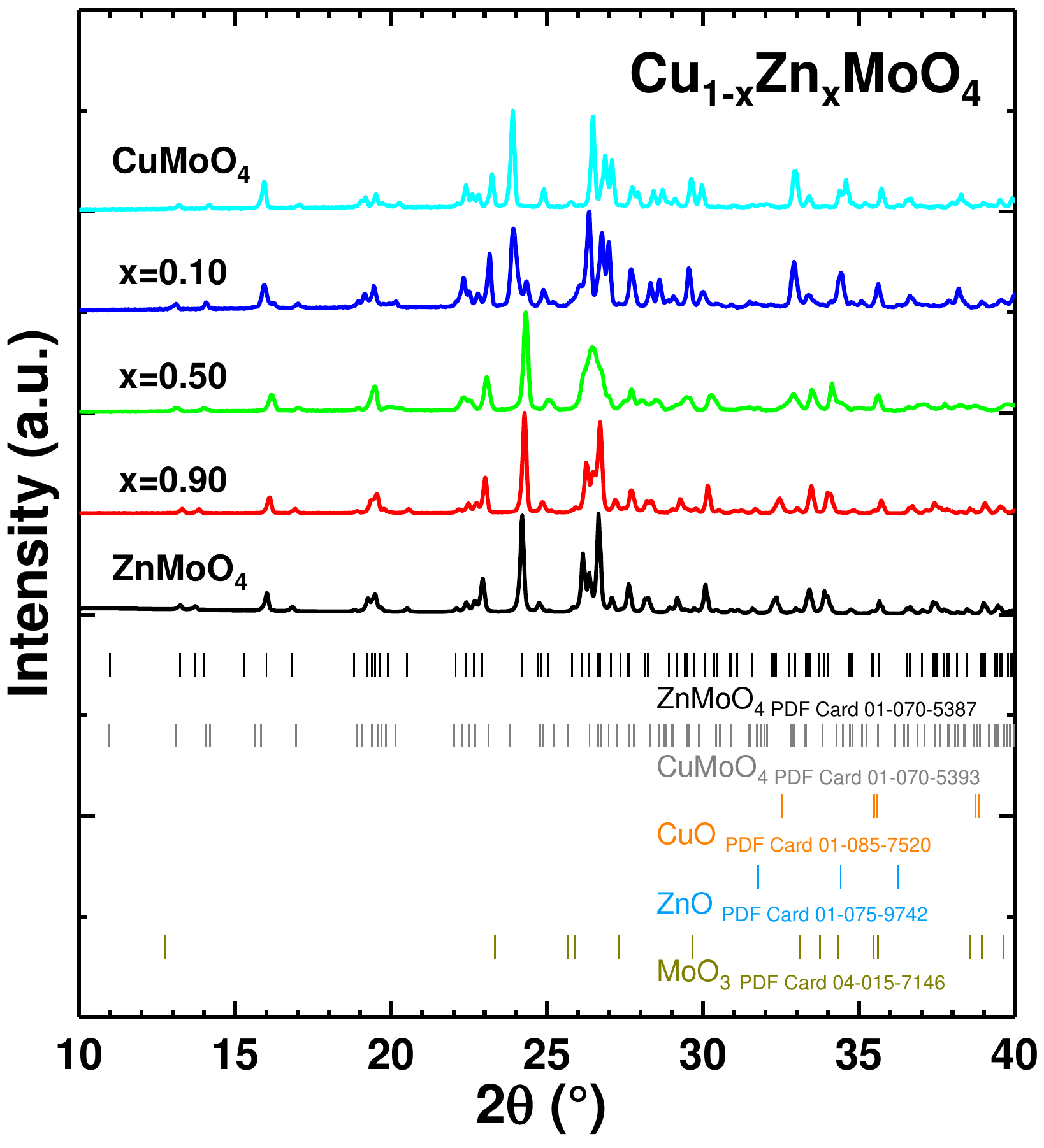}
	\caption{X-ray diffraction patterns of Cu$_{1-x}$Zn$_{x}$MoO$_4$ solid solutions. 
     The standard PDF cards of	ZnMoO$_4$ (PDF Card 01-070-5387), CuMoO$_4$ (PDF Card 01-070-5393), CuO (PDF Card 01-085-7520), ZnO (PDF Card 01-075-9742) and MoO$_3$ 
     (PDF Card 04-015-71446) phases are shown for comparison. }
	\label{fig1}
\end{figure}

The phase purity of all samples was controlled by X-ray diffraction (XRD). 
The XRD patterns (Fig.\ \ref{fig1}) were obtained by a benchtop Rigaku MiniFlex 600 diffractometer with Bragg-Brentano $\theta$-2$\theta$ geometry equipped with the 600~W Cu anode (Cu K$\alpha$ radiation) X-ray tube operated at 40~kV and 15~mA. 
The obtained patterns of solid solutions are in agreement with the literature data \cite{Steiner2003,Reichelt2000}. The contribution of the starting oxide components was not observed which indicates the full completion of the synthesis. 

The X-ray absorption experiments were conducted at HASYLAB PETRA III P65 undulator beamline \cite{P65} as a function of temperature (from 10~K to 300~K) and sample composition. The PETRA III storage ring operated at $E$=6.08~GeV and current $I$=120 mA in top-up 480 bunch mode. The harmonic rejection was achieved by uncoated (Cu/Zn K-edges) and Rh-coated (Mo K-edge) silicon plane mirrors. Fixed exit Si(111) and Si(311) monochromators were used. The X-ray absorption spectra at the Cu (8979~eV), Zn (9659~eV) and Mo (20000~eV) K-edges were collected in transmission mode using two ionisation chambers.  The examined powders were gently milled in an agate mortar and deposited on the Millipore membrane. The temperature-dependent measurements were performed using the Oxford Instruments liquid helium flow cryostat, and the sample temperature was stabilized at each temperature for $\sim$20~min.
The Cu$_{0.90}$Zn$_{0.10}$MoO$_4$ sample was measured on heating from 10~K to 300~K and on cooling from 300~K to 100~K. All other samples were measured only on heating from 10~K to 300~K.

\begin{figure*}[t]
	\centering
	\includegraphics[width=0.95\textwidth]{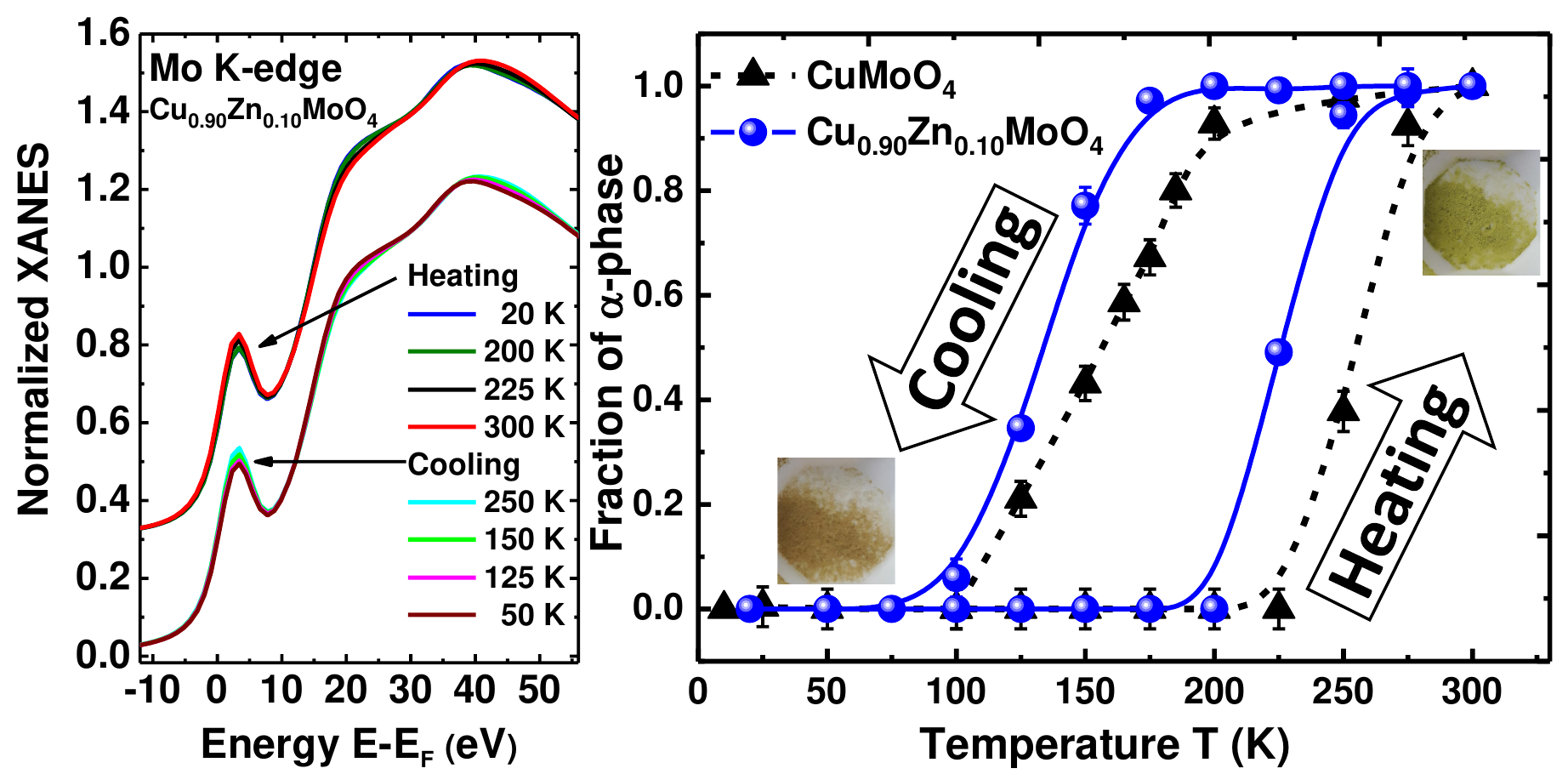}
	\caption{Left panel: Experimental Mo K-edge XANES of Cu$_{0.90}$Zn$_{0.10}$MoO$_4$ as a function of temperature during the sample cooling and heating. Right panel: Temperature dependence of the $\alpha$-phase fraction in CuMoO$_4$ \protect\cite{Jonane2020} and Cu$_{0.90}$Zn$_{0.10}$MoO$_4$ upon heating and cooling. Photographs of the Cu$_{0.90}$Zn$_{0.10}$MoO$_4$ sample at 77~K (brownish) and 300~K (green) are also shown.}
	\label{fig2}
\end{figure*}

Temperature-dependence of the Mo K-edge X-ray absorption near-edge structure (XANES)   in Cu$_{0.90}$Zn$_{0.10}$MoO$_4$ was observed similar to our previous findings for pure CuMoO$_4$ \cite{Jonane2018b}, indicating the change in molybdenum coordination. Therefore, its linear combination analysis (LCA) was performed using one of the low-temperature (50~K) and the highest temperature (300~K) XANES spectra as references. The analysis was done using the Athena package \cite{Ravel2005} in the energy range from 19985~eV to 20015~eV. The obtained results are shown in Fig.\ \ref{fig2}, where the hysteretic behaviour of the fraction of the $\alpha$-phase (300~K) is well observed.

\begin{figure*}[t]
	\centering
	\includegraphics[width=0.7\textwidth]{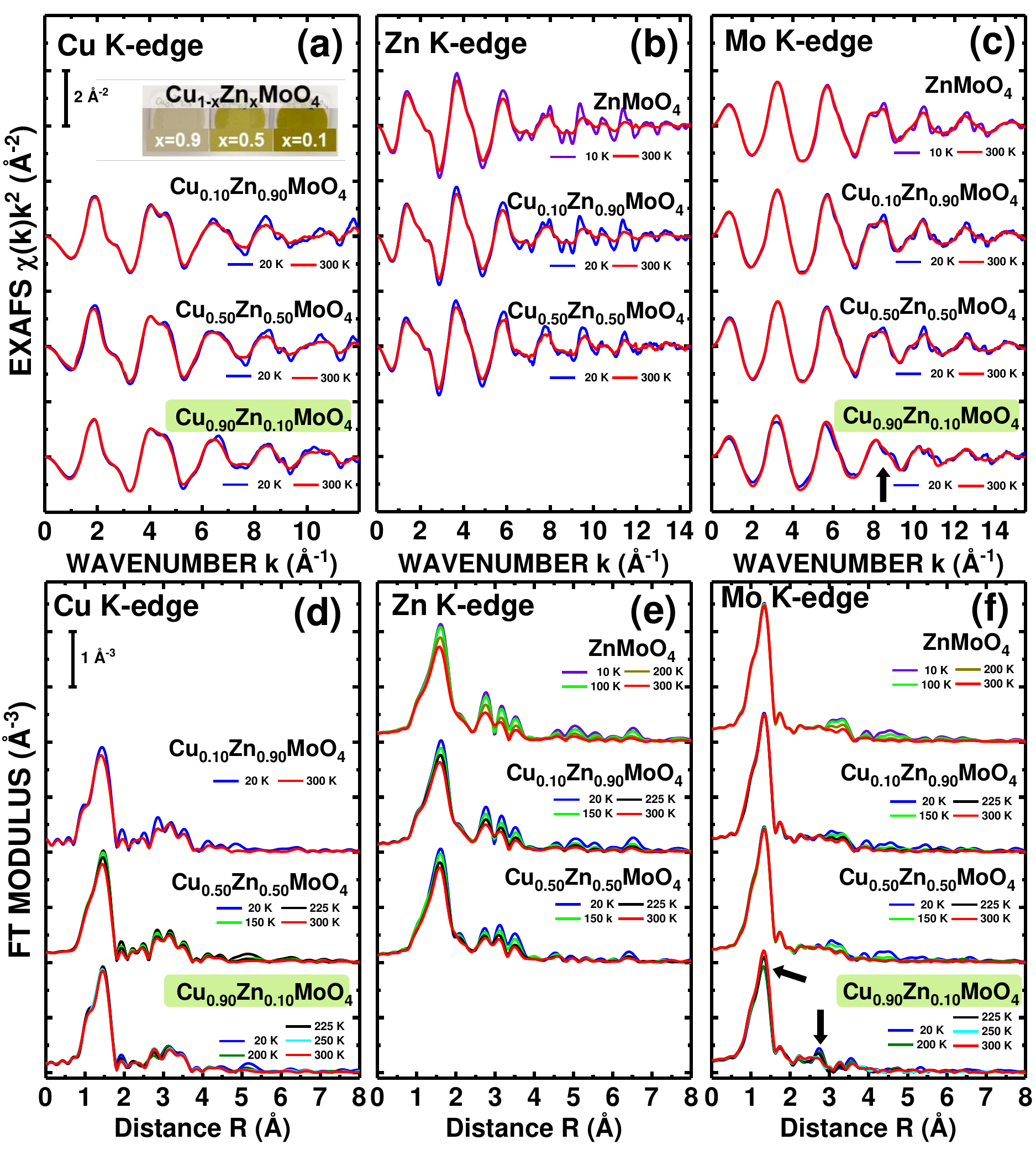}
	\caption{Temperature-dependent experimental Cu, Zn and Mo K-edge EXAFS spectra  $\chi(k)k^2$ for ZnMoO$_4$ and Cu$_{1-x}$Zn$_x$MoO$_4$ ($x$=0.10, 0.50, 0.90) at selected temperatures (a-c) and corresponding Fourier transforms (d-f).  }
	\label{fig3}
\end{figure*}

Temperature-dependent experimental Cu, Zn and Mo K-edge EXAFS $\chi(k)k^2$ spectra of Cu$_{1-x}$Zn$_x$MoO$_4$ were extracted following the conventional
procedure \cite{Kuzmin2014} using the Athena package \cite{Ravel2005}. The EXAFS spectra and their Fourier transforms (FTs) at selected temperatures for several samples are shown in Fig.\ \ref{fig3}. The Zn K-edge EXAFS of Cu$_{0.90}$Zn$_{0.10}$MoO$_4$ was excluded from the analysis due it was corrupted by glitches. The FTs were calculated in the $k$-space range of 1.3–12.0~\AA$^{-1}$\ (for Cu K-edge), 2.3–14.0~\AA$^{-1}$\ (for Zn K-edge) and 2.4–15.0~\AA$^{-1}$\ (for Mo K-edge). The FTs were not corrected for the backscattering phase shift of atoms; therefore, the positions of all peaks are shifted to smaller distances relative to their crystallographic values.

\section{Reverse Monte-Carlo simulations}

The structural information encoded in the experimental EXAFS data was extracted by the reverse Monte-Carlo calculations with an evolutionary algorithm approach (RMC/EA), as implemented in the EvAX code \cite{Timoshenko2014}. The RMC/EA method involves the random changes of atomic coordinates within a three-dimensional structure model of the material, thus minimizing the difference between theoretically calculated configuration-averaged  and experimental EXAFS spectra. The contributions from the structural and thermal disorder in the material, as well as multiple-scattering effects, are taken into account.
Moreover, the method allows one to perform simultaneous fitting of EXAFS data at several absorption edges, which is required to obtain a reliable structure model for such complex materials as solid solutions \cite{Jonane2020,Timoshenko2014analysis,Timoshenko2015}.

Initial structure models for the RMC/EA calculations were constructed based on diffraction data \cite{Wiesmann1997,Abrahams1967,Reichelt2000} by randomly substituting Zn atoms with Cu and vice versa, keeping the stoichiometric ratio of the elements in the supercell (4$a$$\times$4$b$$\times$4$c$) containing 2304 atoms. Several random configurations were simulated. In the case of Cu$_{0.90}$Zn$_{0.10}$MoO$_4$, structure models based on both $\alpha$-CuMoO$_4$ and $\gamma$-CuMoO$_4$ structures were examined, and the relaxed $\alpha$-phase gave a slightly better fit than the $\gamma$-phase for  all examined temperatures. 

\begin{figure*}[t]
	\centering
	\includegraphics[width=0.95\textwidth]{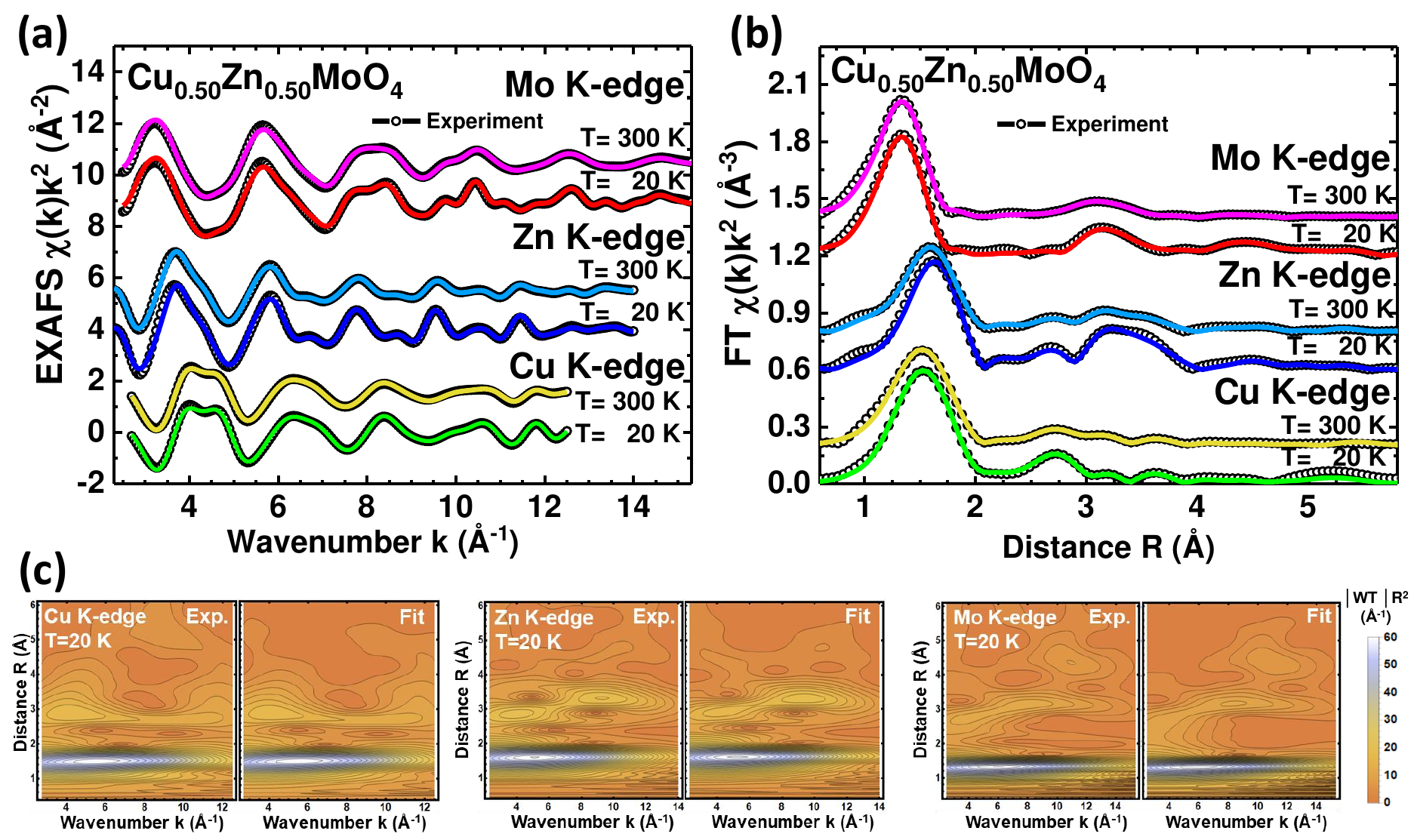}
	\caption{Comparison of the experimental ($T$=20 and 300~K) and calculated configuration-averaged Cu, Zn and Mo K-edge EXAFS spectra of Cu$_{0.50}$Zn$_{0.50}$MoO$_4$ (a) and their Fourier (b) and Morlet wavelet transforms ($T$=20~K) (c).}
	\label{fig4}
\end{figure*}

During RMC/EA simulation, all atoms in the supercell were randomly displaced at each iteration with the maximum allowed displacement of 0.4~\AA. The configuration-averaged EXAFS spectra at the Cu, Zn and Mo K-edges were calculated using the \textit{ab initio} self-consistent real-space multiple-scattering (MS) FEFF8.5L code \cite{Ankudinov1998} taking into account multiple-scattering contributions up to the 5$^{th}$ order. The complex energy-dependent exchange-correlation Hedin-Lundqvist potential was employed to account for inelastic effects \cite{Hedin1971}. The amplitude scaling parameter $S_0^2$ was set to 1. The comparison between the experimental and theoretical EXAFS spectra was carried out in direct ($R$) and reciprocal ($k$) space simultaneously using the Morlet wavelet transform \cite{Timoshenko2009}. The convergence of each RMC simulation was achieved after several thousand iterations. At least three RMC/EA simulations with different sequences of pseudo-random numbers were performed for each experimental data set. The resulting structure models describe well the experimental EXAFS data for all absorption edges (Fig.\ \ref{fig4}).

The final atomic configurations were used to calculate the partial radial distribution functions RDFs $g$($R$) for metal--oxygen and metal--metal atom pairs. 
The widths of RDFs (the variance of interatomic distances) can be described with the mean-square relative displacement factors (MSRDs). They were evaluated directly from atomic coordinates by the median absolute deviation (MAD) method \cite{Daszykowski2007robust}. Information on interatomic interactions was obtained by analysing the temperature dependencies of the MSRD factors $\sigma^2$, which equals to  a sum of thermal $\sigma^2_{th}$ and static $\sigma^2_{st}$ disorder. The temperature dependence of MSRDs was also described by the correlated Einstein model \cite{Sevillano1979}.

\section{Results and discussion}
\label{results}

\subsection{Mo K-edge XANES analysis}

Temperature-dependent normalized Mo K-edge XANES spectra of Cu$_{0.90}$Zn$_{0.10}$MoO$_4$ are shown in Fig.\ \ref{fig2}. 
The pre-peak just above the Fermi level corresponds to the 1s(Mo)$\rightarrow$4d(Mo)+2p(O) transition. The linear combination analysis (LCA) of the Mo K-edge XANES was performed in the temperature range from 20 to 300~K. In analogy with pure CuMoO$_4$ \cite{Jonane2018b,Jonane2020}, we assume that Cu$_{0.90}$Zn$_{0.10}$MoO$_4$ exists in $\alpha$-phase at 300~K, while in $\gamma$-phase  at 20~K. Note that the sample colour changes from green to brownish after treatment in liquid nitrogen ($\sim$77~K) and returns to green after heating up to 300~K as shown in Fig.\ \ref{fig2}. 

A fraction of the $\alpha$-phase in Cu$_{0.90}$Zn$_{0.10}$MoO$_4$ at different temperatures is shown in Fig.\ \ref{fig2} and demonstrates hysteresis behavior on cooling and heating. 
Notably that the phase transition hysteresis in Cu$_{0.90}$Zn$_{0.10}$MoO$_4$ is shifted to lower temperatures compared to pure CuMoO$_4$ by about 25~K. At the same time, the variation of the pre-edge peak for Cu$_{0.90}$Zn$_{0.10}$MoO$_4$ is not as pronounced as for CuMoO$_4$ reported in \cite{Jonane2018b} indicating on slightly different molybdenum local environment at low temperatures. Based on XANES data, the transition from the $\alpha$ to $\gamma$-phase occurs in Cu$_{0.90}$Zn$_{0.10}$MoO$_4$ between $\sim$175~K and $\sim$100~K upon cooling, whereas the transition from the $\gamma$ to $\alpha$ phase begins above $\sim$200~K and ends at $\sim$250~K.

\subsection{Experimental EXAFS data}

Experimental Cu, Zn and Mo K-edge EXAFS spectra $\chi(k)k^2$  of Cu$_{1-x}$Zn$_x$MoO$_4$ together with their FTs are compared in Fig.\ \ref{fig3}. The first peak located at about 1-2~\AA\ in FTs is due to the contribution from the first coordination shell formed by oxygen atoms. More distant FT peaks originate from single-scattering contributions of outer coordination shells and multiple-scattering effects. The amplitude of the FT peaks changes upon temperature variation due to the thermal disorder effect. However, contributions up to $\sim$6-7~\AA\  can be observed even at 300~K. 
 
While the temperature-dependence at the Mo K-edge (especially at low-$k$ values) for samples with $x$$\geq$0.50 is very weak, significant changes occur in the first  coordination shell of molybdenum in Cu$_{0.90}$Zn$_{0.10}$MoO$_4$ sample. 
Indeed, the amplitude of the FT peak at 1.3~\AA\ increases slightly on heating 
suggesting the less distorted local environment of molybdenum at 300~K (Fig.\ \ref{fig3}(f)). 
Such behaviour is similar to the molybdenum change from octahedral to tetrahedral coordination during the $\gamma$-to-$\alpha$ phase transition in pure CuMoO$_4$ \cite{Jonane2018c}. 
 
Comparing EXAFS spectra at the Cu and Zn K-edges, one can see that the local environment of zinc atoms exhibits a stronger temperature dependence than that of copper atoms. The damping of the Zn K-edge EXAFS oscillations increases noticeably at high-$k$ values with temperature increase due to enhanced thermal disorder.

\subsection{Results of RMC/EA simulations}
 
Detailed structural models were extracted from the experimental EXAFS spectra using 
RMC/EA calculations taking into account simultaneously two (in ZnMoO$_4$ and Cu$_{0.90}$Zn$_{0.10}$MoO$_4$) or three (in Cu$_{1-x}$Zn$_{x}$MoO$_4$ with $x$=0.50 and 0.90) metal absorption edges. An example of the obtained EXAFS fits, along with corresponding Fourier and Wavelet transforms of the experimental and theoretical EXAFS spectra, are shown in Fig.\ \ref{fig4} on the example of Cu$_{0.50}$Zn$_{0.50}$MoO$_4$ sample. Good agreement between the experimental and simulated data was achieved for all three metal absorption edges at all temperatures. Note that the contribution from the first coordination shell of metal atoms dominates EXAFS, FT and WT signals, however, the outer shells are also detectable up to 5-6~\AA.  

\begin{figure}[t]
	\centering
	\includegraphics[width=0.55\textwidth]{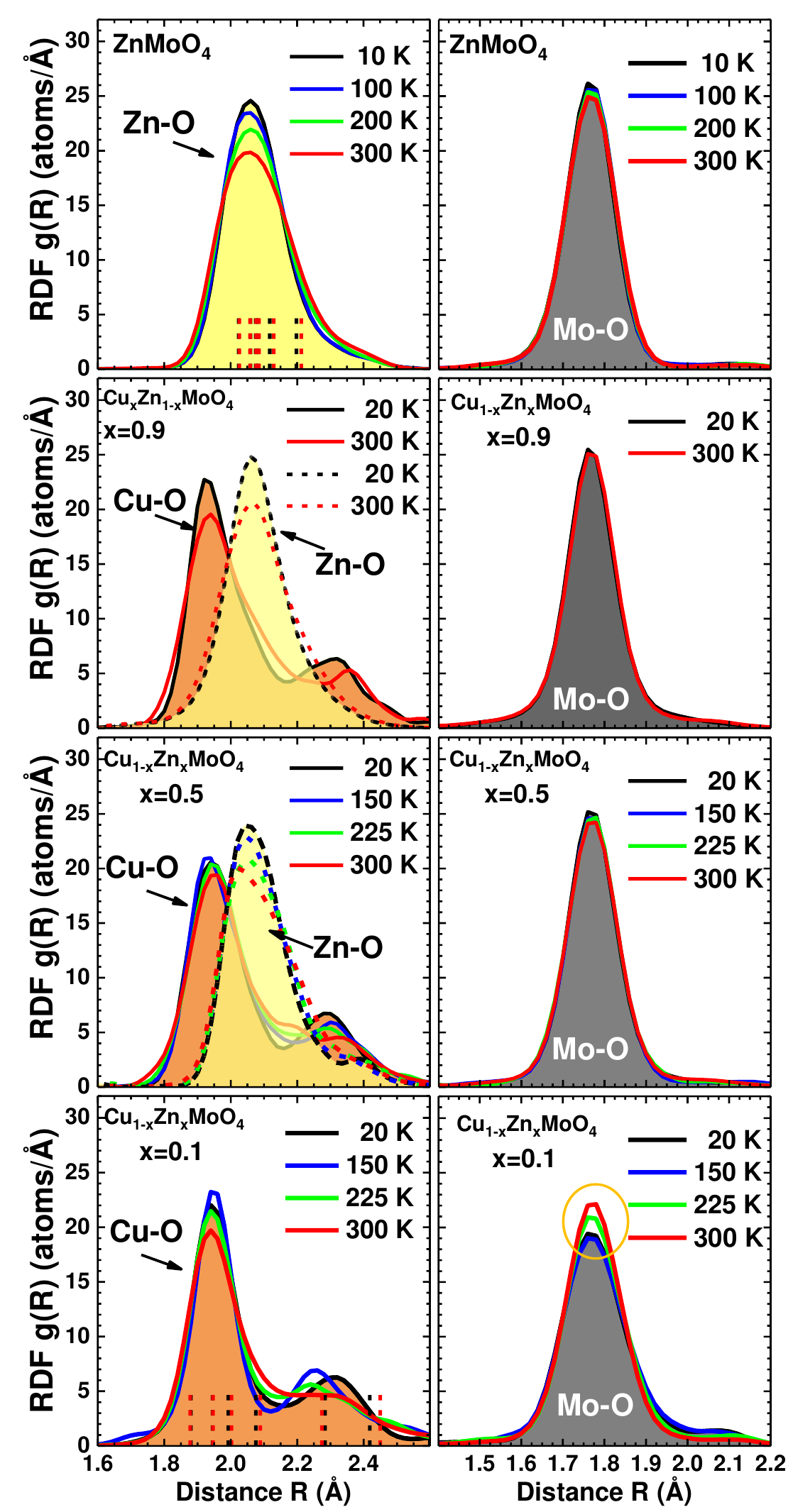}
	\caption{Partial RDFs around zinc and molybdenum atoms in ZnMoO$_4$ at $T$=10~K and 300~K obtained in RMC/EA simulations. Contributions from different atom pairs are indicated.}
	\label{fig5}
\end{figure}

Final sets of atomic coordinates obtained in the RMC/EA simulations were used to calculate partial radial distribution functions (RDFs) around absorbing atoms. They  are reported up to 6~\AA\ for ZnMoO$_4$ at 10~K and 300~K in Fig.\ \ref{fig5}.
As one can see, the first coordination shell is well separated, while outer shells overlap strongly. Therefore, we will concentrate further on the analysis and discussion of the first coordination shell only.

\begin{figure}[t]
	\centering
	\includegraphics[width=0.55\textwidth]{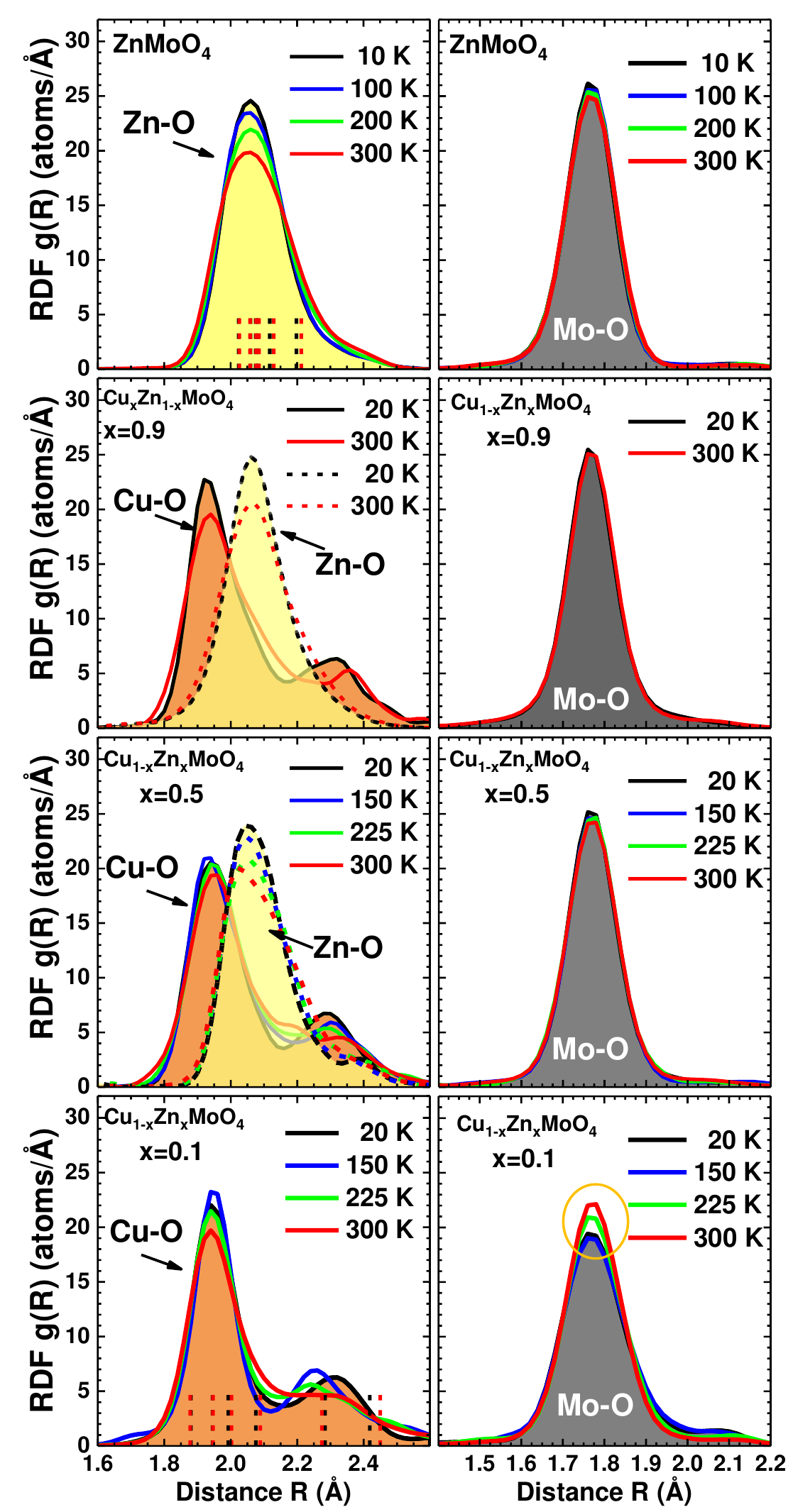}
	\caption{Temperature-dependent partial RDFs $g_{\rm  Zn-O}$($R$), $g_{\rm Cu-O}$($R$) and $g_{\rm Mo-O}$($R$) obtained in RMC/EA simulations. Vertical dashed lines in $g_{\rm Zn-O}$($R$) plot for ZnMoO$_4$  and $g_{\rm Cu-O}$($R$) plot for Cu$_{0.90}$Zn$_{0.10}$MoO$_4$ indicate average Zn--O and  Cu--O interatomic distances, respectively, at 10~K(20~K) and 300~K (black and red lines, respectively). }
	\label{fig6}
\end{figure}

Temperature dependence of partial RDFs $g_{\rm Zn-O}$($R$), $g_{\rm Cu-O}$($R$),  and $g_{\rm Mo-O}$($R$) for ZnMoO$_4$ and Cu$_{1-x}$Zn$_x$MoO$_4$ is shown in Fig.\ \ref{fig6}.
  
The RDFs $g_{\rm Cu-O}$($R$) for Cu--O atom pairs have two maxima at $\sim$1.9~\AA\ and $\sim$2.3~\AA\ with a relative area of about 4:2. Strong distortion of the first coordination shell of copper ions is caused by the first-order Jahn-Teller effect due to the Cu$^{2+}$[3d$^9$] electronic configuration. The distribution of the nearest four oxygen atoms, located in the plane of the CuO$_6$ octahedron/CuO$_5$ square-pyramid, has a rather weak dependence on temperature due to strong Cu--O bonds. The two axial oxygen atoms contribute to the second broad peak of the RDF. Averaged  Zn--O and  Cu--O interatomic distances calculated from atomic coordinates in the structural models are shown with vertical dashed lines in Fig.\ \ref{fig6} for ZnMoO$_4$  and  Cu$_{0.90}$Zn$_{0.10}$MoO$_4$, respectively. One can see that during heating, the longest Cu--O bond  slightly lengthens. 
 
The local environment of zinc ions differs from that of copper, since Zn$^{2+}$ has an electronic configuration 3d$^{10}$ and, thus, the Jahn-Teller distortion is absent. The distribution $g_{\rm Zn-O}$($R$) can be characterized with one asymmetric peak with a maximum at about 2.06~\AA, due to both ZnO$_6$ octahedron/ZnO$_5$ square-pyramid are slightly deformed. 

The distribution $g_{\rm Mo-O}$($R$) of the nearest four oxygen atoms, forming MoO$_4$ tetrahedron with $R$(Mo--O)$\approx$1.77~\AA, is rather narrow and has a weak dependence on temperature due to strong Mo--O bonds. The RDFs $g_{\rm Mo-O}$($R$) in pure $\alpha$-ZnMoO$_4$ and Cu$_{1-x}$Zn$_x$MoO$_4$ solid solutions with $x$=0.50 and 0.90 remain nearly unchanged in the temperature range of 20-300~K. 
At the same time, some variation of the $g_{\rm Mo-O}$($R$) upon temperature increase 
is observed in Cu$_{0.90}$Zn$_{0.10}$MoO$_4$ and related to structural changes.

\subsection{MSRD factors}
 
The widths of RDFs can be characterized by the MSRD factors $\sigma^2$. 
However, due to the distorted environment of metal ions, only the groups of nearest oxygen atoms, which are responsible for the main peak in partial RDFs in Fig.\ \ref{fig6}, have been considered in the evaluation of $\sigma^2$. 

\begin{figure*}[t]
	\centering
	\includegraphics[width=0.9\textwidth]{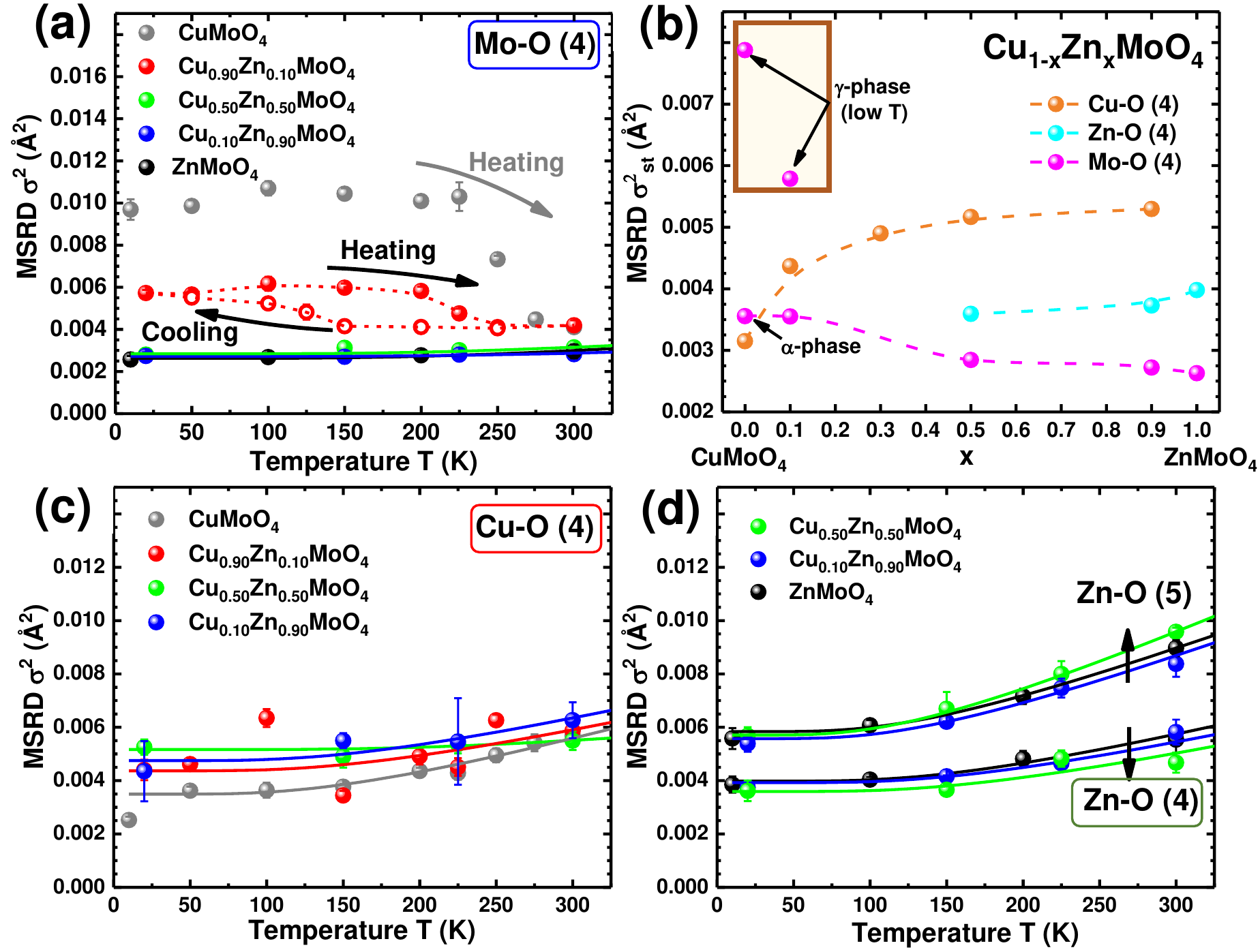}
	\caption{Temperature dependencies of the MSRD factors $\sigma^2$, obtained in RMC/EA simulations for metal (Mo, Cu, Zn) and surrounding four nearest oxygen atoms (a, c, d). Additionally MSRD $\sigma^2$ calculated for zinc and five nearest oxygens is plotted in panel (d) for comparison. Static disorder term $\sigma^2_{st}$ as a function of zinc concentration $x$ in Cu$_{1-x}$Zn$_x$MoO$_4$ is shown in panel (b). Solid lines are fits by the correlated Einstein model. Dashed lines are guide for eye. }
	\label{fig7}
\end{figure*}

The calculated  MSRDs are plotted as a function of temperature and composition in Fig.\ \ref{fig7}. The temperature dependencies were fitted with the correlated Einstein model \cite{Sevillano1979}, and the static disorder factors $\sigma_{st}$ were estimated.

Similar to the result of the Mo K-edge XANES analysis (Fig.\ \ref{fig2}), the temperature-dependence of Mo--O distances in Cu$_{0.90}$Zn$_{0.10}$MoO$_4$ shows hysteretic behaviour (Fig.\ \ref{fig7}(a)). Note that the MSRD values $\sigma^2$(Mo--O) at temperatures below 250~K are much smaller than for isostructural low-temperature  $\gamma$-CuMoO$_4$. At the same time, the value of the MSRD factor at 300~K is close to that in high-temperature $\alpha$-CuMoO$_4$.  

No hysteresis was found for pure ZnMoO$_4$ and Cu$_{1-x}$Zn$_x$MoO$_4$ solid solutions with $x$=0.50 and 0.90. In fact, small values and weak temperature dependence of their MSRDs indicate strong Mo--O bonding, which is characteristic of MoO$_4$ tetrahedral coordination. The estimated Einstein frequency is equal to about $\omega_E$=111$\pm$52~THz.

The MSRDs values  $\sigma^2$(Cu--O) for the four shortest Cu--O bonds are close in all Cu$_{1-x}$Zn$_x$MoO$_4$ solid solutions and CuMoO$_4$ from \cite{Jonane2020} within error bars (Fig.\ \ref{fig7}(c)). Their Einstein frequencies are in the range of  $\omega_E$=78$\pm$39~THz. 

The MSRD  values $\sigma^2$(Zn--O) for groups of the nearest four (Zn--O(4)) or five (Zn--O(5)) oxygen atoms show weak dependence on the composition (Fig.\ \ref{fig7}(d)) being close in $\alpha$-ZnMoO$_4$ and Cu$_{1-x}$Zn$_{x}$MoO$_4$ samples with $x$=0.90 and 0.50. The calculated Einstein frequencies are equal to $\omega_E$=68$\pm$10~THz when one considers only four nearest oxygen atoms and $\omega_E$=52$\pm$4~THz when the fifth oxygen atom is additionally included.

The composition dependence of static distortions around metal ions in Cu$_{1-x}$Zn$_x$MoO$_4$ solid solutions can be evaluated based on the values of $\sigma^2_{st}$ for the nearest four Cu--O, Zn--O and Mo--O atom pairs (Fig.\ \ref{fig7}(b)). 

The $\sigma^2_{st}$ values for the Mo--O bonds reveal that the static distortions are much larger in the low-temperature $\gamma$-CuMoO$_4$ than in the low-temperature phase of Cu$_{0.90}$Zn$_{0.10}$MoO$_4$, both having distorted octahedral coordination of molybdenum ions.

Note that the 50\% of $\alpha$-Cu$_{0.90}$Zn$_{0.10}$MoO$_4$ transforms to the low-temperature phase during cooling at $T_{1/2C}\approx$125~K (134~K from XANES in Fig.\ \ref{fig2}) and returns on heating at $T_{1/2H}\approx$225~K  (Fig.\ \ref{fig7}(a)).
Both $T_{1/2C}$ and  $T_{1/2H}$ values are shifted to lower temperatures in comparison to pure CuMoO$_4$, where $T_{1/2C}\approx$143-155~K and $T_{1/2H}\approx$255~K (Fig.\ \ref{fig2} and \cite{Jonane2018b}). Thus, the substitution of Cu$^{2+}$ ions with Zn$^{2+}$ ions stabilizes the $\alpha$-phase, which is natural for $\alpha$-ZnMoO$_4$.

The tetrahedral coordination of molybdenum ions at room-temperature in $\alpha$-CuMoO$_4$, 
$\alpha$-ZnMoO$_4$ and $\alpha$-Cu$_{1-x}$Zn$_{x}$MoO$_4$ solid solutions is responsible for small values of $\sigma^2_{st}$ for Mo--O bonds in Fig.\ \ref{fig7}(b).
The distorted octahedral environment of zinc ions results in larger values of $\sigma^2_{st}$ for the Zn--O bonds and is almost constant at large $x$$\geq$0.5. 
An increase of $\sigma^2_{st}$ for the Cu--O bonds can be related to the role played by copper ions upon increasing $x$. Static distortions are smallest in pure $\alpha$-CuMoO$_4$, increase upon a formation of the solid solution with a maximum at $x$=0.50 and remain constant for a small amount of copper ($x$$>$0.50) when the structure is determined by the zinc sublattice. 
 
To conclude, the local environment of molybdenum ions in Cu$_{1-x}$Zn$_{x}$MoO$_4$ solid solutions can be significantly affected by the substitution of copper with zinc, since the electronic configurations of the two 3d-ions differ significantly. Indeed, the 3d(Zn$^{2+}$)-subshell is filled, while the 3d$^9$ configuration of Cu$^{2+}$ makes it subject to the first-order Jahn-Teller distortion leading to tetragonal deformation of CuO$_6$ octahedra.  The substitution modifies the electronic structure of molybdate by increasing its band gap \cite{Tiwari2020} and leads to a  disappearance of thermochromic properties, which are closely related to the local structure of molybdenum ions.
The coordination of  molybdenum ions changes from distorted octahedral at low-temperature in $\gamma$-Cu$_{0.90}$Zn$_{0.10}$MoO$_4$ solid solution to tetrahedral in $\alpha$-Cu$_{0.90}$Zn$_{0.10}$MoO$_4$ above 250~K,
similar to temperature or pressure-induced the $\gamma$-to-$\alpha$ phase transitions in CuMoO$_4$ \cite{Wiesmann1997}. At the same time, molybdenum ions have always 
tetrahedral coordination in $\alpha$-Cu$_{1-x}$Zn$_{x}$MoO$_4$  with $x$$>$0.10 and $\alpha$-ZnMoO$_4$.

\section{Conclusions}
\label{conc}

The influence of zinc ions on the thermochromic properties of polycrystalline Cu$_{1-x}$Zn$_x$MoO$_4$ ($x$=0.10, 0.50, 0.90) solid solutions was studied by X-ray absorption spectroscopy  at the Cu, Zn and Mo K-edges. The analysis of XANES and EXAFS spectra revealed that the substitution of Cu$^{2+}$ ions with Zn$^{2+}$ ions stabilizes the $\alpha$-phase, which is natural for $\alpha$-ZnMoO$_4$, and  significantly affects the temperature of the thermochromic phase transition in the materials.

We found that among sampled solid solutions, only Cu$_{0.90}$Zn$_{0.10}$MoO$_4$ exhibited a  thermochromic phase transition between $\alpha$ and $\gamma$ phases with the hysteretic behaviour. This result is in agreement with that of the previous work \cite{Asano2011}.

To understand the structural origin of the thermochromic phase transition in Cu$_{0.90}$Zn$_{0.10}$MoO$_4$, a detailed analysis of the mean-square relative displacement factors for metal--oxygen bonds in the first coordination shell was performed. We found that the substitution of copper with zinc affects strongly the local environment of molybdenum ions. This result can be explained by the instability of molybdenum coordination, which is tetrahedral in $\alpha$-Cu$_{0.90}$Zn$_{0.10}$MoO$_4$ solid solution at room temperature
but changes to distorted octahedral under lattice contraction at low-temperature in $\gamma$-Cu$_{0.90}$Zn$_{0.10}$MoO$_4$, similar to 
temperature or pressure-induced the $\alpha$-to-$\gamma$ phase transitions in CuMoO$_4$ 
\cite{Wiesmann1997}.

Thus, tuning the thermochromic properties of CuMoO$_4$ by doping with zinc may find useful applications for controlling storage conditions in the low-temperature range.

\section*{Acknowledgements}

I. P., A. K. and A. K. would like to thank the support of the Latvian Council of Science project No. lzp-2019/1-0071. 
I.P. acknowledges the L‘ORÉAL Baltic “For Women In Science” Program with the support of the Latvian National Commission for UNESCO and the Latvian Academy of Sciences.
The experiment at the PETRA III synchrotron was performed within project No. I-20190277 EC. The synchrotron experiments have been supported by the project CALIPSOplus under the Grant Agreement 730872 from the EU Framework Programme for Research and Innovation HORIZON 2020. 
Institute of Solid State Physics, University of Latvia as the Center of Excellence has received funding from the European Union’s Horizon 2020 Framework Programme H2020-WIDESPREAD-01-2016-2017-TeamingPhase2 under grant agreement No. 739508, project CAMART2.

\section*{Data availability}
The raw/processed data required to reproduce these findings cannot be shared at this
time due to technical or time limitations.


\begin{thebibliography}{10}
	\expandafter\ifx\csname url\endcsname\relax
	\def\url#1{\texttt{#1}}\fi
	\expandafter\ifx\csname urlprefix\endcsname\relax\def\urlprefix{URL }\fi
	\expandafter\ifx\csname href\endcsname\relax
	\def\href#1#2{#2} \def\path#1{#1}\fi
	
	\bibitem{Day1968}
	J.~H. Day, {Thermochromism of inorganic compounds}, Chem. Rev. 68 (1968)
	649--657.
	\newblock \href {https://doi.org/10.1021/cr60256a001}
	{\path{doi:10.1021/cr60256a001}}.
	
	\bibitem{Kiri2010}
	P.~Kiri, R.~B. G.~Hyett, {Solid state thermochromic materials}, Adv. Mat. Lett.
	1 (2010) 86--105.
	\newblock \href {https://doi.org/10.5185/amlett.2010.8147}
	{\path{doi:10.5185/amlett.2010.8147}}.
	
	\bibitem{Wang2016}
	Y.~Wang, E.~L. Runnerstom, D.~J. Milliron, {Switchable materials for smart
		windows}, Annu. Rev. Chem. Biomol. Eng. 7 (2016) 283--304.
	\newblock \href {https://doi.org/10.1146/annurev-chembioeng-080615-034647}
	{\path{doi:10.1146/annurev-chembioeng-080615-034647}}.
	
	\bibitem{Steiner2001}
	G.~Steiner, R.~Salzer, W.~Reichelt, {Temperature dependence of the optical
		properties of CuMoO$_4$}, Fresenius J. Anal. Chem. 370 (2001) 731--734.
	\newblock \href {https://doi.org/10.1007/s002160000630}
	{\path{doi:10.1007/s002160000630}}.
	
	\bibitem{Gaudon2007a}
	M.~Gaudon, C.~Carbonera, A.~E. Thiry, A.~Demourgues, P.~Deniard, C.~Payen,
	J.~F. L\'etard, S.~Jobic, {Adaptable thermochromism in the
		CuMo$_{1-x}$W$_x$O$_4$ series ($0 \leq x < 0.1$): A behavior related to a
		first-order phase transition with a transition temperature depending on x},
	Inorg. Chem. 46 (2007) 10200--10207.
	\newblock \href {https://doi.org/10.1021/ic701263c}
	{\path{doi:10.1021/ic701263c}}.
	
	\bibitem{Jonane2019b}
	I.~Jonane, A.~Anspoks, G.~Aquilanti, A.~Kuzmin, {High-temperature X-ray
		absorption spectroscopy study of thermochromic copper molybdate}, Acta Mater.
	179 (2019) 26--35.
	\newblock \href {https://doi.org/10.1016/j.actamat.2019.06.034}
	{\path{doi:10.1016/j.actamat.2019.06.034}}.
	
	\bibitem{Joseph2020}
	N.~Joseph, J.~Varghese, M.~Teirikangas, H.~Jantunen, {A temperature-responsive
		copper molybdate polymorph mixture near to water boiling point by a simple
		cryogenic quenching route}, ACS Appl. Mater. Interfaces 12 (2020) 1046--1053.
	\newblock \href {https://doi.org/10.1021/acsami.9b17300}
	{\path{doi:10.1021/acsami.9b17300}}.
	
	\bibitem{Gaudon2007b}
	M.~Gaudon, P.~Deniard, A.~Demourgues, A.~E. Thiry, C.~Carbonera, A.~{Le
		Nestour}, A.~Largeteau, J.~F. L\'etard, S.~Jobic, {Unprecedented
		``one-finger-push''-induced phase transition with a drastic color change in
		an inorganic material}, Adv. Mater. 19 (2007) 3517--3519.
	\newblock \href {https://doi.org/10.1002/adma.200700905}
	{\path{doi:10.1002/adma.200700905}}.
	
	\bibitem{Gaudon2010}
	M.~Gaudon, C.~Riml, A.~Turpain, C.~Labrugere, M.~H. Delville, {Investigation of
		the chromic phase transition of CuMo$_{0.9}$W$_{0.1}$O$_4$ induced by surface
		protonation}, Chem. Mater. 22 (2010) 5905--5911.
	\newblock \href {https://doi.org/10.1021/cm101824d}
	{\path{doi:10.1021/cm101824d}}.
	
	\bibitem{Yanase2013}
	I.~Yanase, T.~Mizuno, H.~Kobayashi, {Structural phase transition and
		thermochromic behavior of synthesized W-substituted CuMoO$_4$}, Ceram. Int.
	39 (2013) 2059--2064.
	\newblock \href {https://doi.org/10.1016/j.ceramint.2012.08.059}
	{\path{doi:10.1016/j.ceramint.2012.08.059}}.
	
	\bibitem{Blanco2015}
	V.~Blanco-Gutierrez, L.~Cornu, A.~Demourgues, M.~Gaudon,
	{CoMoO$_4$/CuMo$_{0.9}$W$_{0.1}$O$_4$} mixture as an efficient piezochromic
	sensor to detect temperature/pressure shock parameters, ACS Appl. Mater.
	Interfaces 7 (2015) 7112--7117.
	\newblock \href {https://doi.org/10.1021/am508652h}
	{\path{doi:10.1021/am508652h}}.
	
	\bibitem{Robertson2015}
	L.~Robertson, N.~Penin, V.~Blanco-Gutierrez, D.~Sheptyakov, A.~Demourgues,
	M.~Gaudon, {CuMo$_{0.9}$W$_{0.1}$O$_4$} phase transition with thermochromic,
	piezochromic, and thermosalient effects, J. Mater. Chem. C 3 (2015)
	2918--2924.
	\newblock \href {https://doi.org/10.1039/C4TC02463J}
	{\path{doi:10.1039/C4TC02463J}}.
	
	\bibitem{Yanase2019}
	I.~Yanase, R.~Koda, R.~Kondo, R.~Taiji, {Improvement of thermochromic property
		at low temperatures of CuMo$_{0.94}$W$_{0.06}$O$_4$ by Zn substitution}, J.
	Therm. Anal. Calorim. 140 (2019) 2203--2214.
	\newblock \href {https://doi.org/10.1007/s10973-019-09025-7}
	{\path{doi:10.1007/s10973-019-09025-7}}.
	
	\bibitem{DaSilva2020}
	M.~V. {da Silva}, D.~F.~M. {de Oliveira}, H.~S. Oliveira, K.~P.~F. Siqueira,
	{Influence of temperature on the structural and color properties of nickel
		molybdates}, Mater. Res. Bull. 122 (2020) 110665.
	\newblock \href {https://doi.org/10.1016/j.materresbull.2019.110665}
	{\path{doi:10.1016/j.materresbull.2019.110665}}.
	
	\bibitem{Costa2020}
	R.~K.~S. Costa, S.~C. Teles, K.~P.~F. Siqueira, {The relationship between
		crystal structures and thermochromism in CoMoO$_4$}, Chem. Pap. 75 (2020)
	237--248.
	\newblock \href {https://doi.org/10.1007/s11696-020-01294-z}
	{\path{doi:10.1007/s11696-020-01294-z}}.
	
	\bibitem{Wiesmann1997}
	M.~Wiesmann, H.~Ehrenberg, G.~Miehe, T.~Peun, H.~Weitzel, H.~Fuess, {$p-T$
		phase diagram of CuMoO$_4$}, J. Solid State Chem. 132 (1997) 88--97.
	\newblock \href {https://doi.org/10.1006/jssc.1997.7413}
	{\path{doi:10.1006/jssc.1997.7413}}.
	
	\bibitem{Jonane2018b}
	I.~Jonane, A.~Cintins, A.~Kalinko, R.~Chernikov, A.~Kuzmin, {X-ray absorption
		near edge spectroscopy of thermochromic phase transition in CuMoO$_4$}, Low
	Temp. Phys. 44 (2018) 434--437.
	\newblock \href {https://doi.org/10.1063/1.5034155}
	{\path{doi:10.1063/1.5034155}}.
	
	\bibitem{Abrahams1967}
	S.~Abrahams, {Crystal structure of the transition-metal molybdates and
		tungstates. III. Diamagnetic $\alpha$-ZnMoO$_4$}, J. Chem. Phys. 46 (1967)
	2052--2063.
	\newblock \href {https://doi.org/10.1063/1.1841001}
	{\path{doi:10.1063/1.1841001}}.
	
	\bibitem{Yadav2019}
	P.~Yadav, E.~Sinha, {Structural, photophysical and microwave dielectric
		properties of $\alpha$-ZnMoO$_4$ phosphor}, J. Alloys Compd. 795 (2019)
	446--452.
	\newblock \href {https://doi.org/10.1016/j.jallcom.2019.05.019}
	{\path{doi:10.1016/j.jallcom.2019.05.019}}.
	
	\bibitem{Steiner2003}
	U.~Steiner, W.~Reichelt, S.~D{\"a}britz, {Chemischer Transport von
		Mischkristallen im System CuMoO$_4$/ZnMoO$_4$}, Z. Anorg. Allg. Chem. 629
	(2003) 116--122.
	\newblock \href {https://doi.org/10.1002/zaac.200390002}
	{\path{doi:10.1002/zaac.200390002}}.
	
	\bibitem{Cavalcante2013}
	L.~Cavalcante, E.~Moraes, M.~Almeida, C.~Dalmaschio, N.~Batista, J.~Varela,
	E.~Longo, M.~{Siu Li}, J.~Andres, A.~Beltr\'an, {A combined theoretical and
		experimental study of electronic structure and optical properties of
		$\beta$-ZnMoO$_4$ microcrystals}, Polyhedron 54 (2013) 13--25.
	\newblock \href {https://doi.org/10.1016/j.poly.2013.02.006}
	{\path{doi:10.1016/j.poly.2013.02.006}}.
	
	\bibitem{MtiouiSghaier2015}
	O.~Mtioui-Sghaier, R.~Mendoza-Mero{\~{n}}o, L.~Ktari, M.~Dammak,
	S.~Garc{\'\i}a-Granda, {Redetermination of the crystal structure of
		{$\beta$}-zinc molybdate from single-crystal X-ray diffraction data}, Acta
	Crystallogr. E 71 (2015) i6--i7.
	\newblock \href {https://doi.org/10.1107/S205698901501186X}
	{\path{doi:10.1107/S205698901501186X}}.
	
	\bibitem{Reichelt2000}
	W.~Reichelt, T.~Weber, T.~S{\"o}hnel, S.~D{\"a}britz, {Mischkristallbildung im
		System CuMoO$_4$/ZnMoO$_4$}, Z. Anorg. Allg. Chem. 626 (2000) 2020--2027.
	\newblock \href
	{https://doi.org/10.1002/1521-3749(200009)626:9<2020::AID-ZAAC2020>3.0.CO;2-K}
	{\path{doi:10.1002/1521-3749(200009)626:9<2020::AID-ZAAC2020>3.0.CO;2-K}}.
	
	\bibitem{Tiwari2020}
	S.~K. Tiwari, A.~Singh, P.~Yadav, B.~K. Sonu, R.~Verma, S.~Rout, E.~Sinha,
	{Structural and dielectric properties of Cu-doped $\alpha$-ZnMoO$_4$ ceramic
		system for enhanced green light emission and potential microwave
		applications}, J. Mater. Sci.: Mater. Electron. (2020).
	\newblock \href {https://doi.org/10.1007/s10854-020-04225-6}
	{\path{doi:10.1007/s10854-020-04225-6}}.
	
	\bibitem{Asano2011}
	T.~Asano, T.~Nishimura, S.~Ichimura, Y.~Inagaki, T.~Kawae, T.~Fukui, Y.~Narumi,
	K.~Kindo, T.~Ito, S.~Haravifard, et~al., {Magnetic ordering and tunable
		structural phase transition in the chromic compound CuMoO$_4$}, J. Phys. Soc.
	Jpn. 80 (2011) 093708.
	\newblock \href {https://doi.org/10.1143/JPSJ.80.093708}
	{\path{doi:10.1143/JPSJ.80.093708}}.
	
	\bibitem{Timoshenko2014}
	J.~Timoshenko, A.~Kuzmin, J.~Purans, {EXAFS study of hydrogen intercalation
		into ReO$_3$ using the evolutionary algorithm}, J. Phys.: Condens. Matter 26
	(2014) 055401.
	\newblock \href {https://doi.org/10.1088/0953-8984/26/5/055401}
	{\path{doi:10.1088/0953-8984/26/5/055401}}.
	
	\bibitem{Jonane2020}
	I.~Jonane, A.~Cintins, A.~Kalinko, R.~Chernikov, A.~Kuzmin, {Low temperature
		X-ray absorption spectroscopy study of CuMoO$_4$ and
		CuMo$_{0.90}$W$_{0.10}$O$_4$ using reverse Monte-Carlo method}, Rad. Phys.
	Chem. 175 (2020) 108411.
	\newblock \href {https://doi.org/0.1016/j.radphyschem.2019.108411}
	{\path{doi:0.1016/j.radphyschem.2019.108411}}.
	
	\bibitem{Jonane2018c}
	I.~Jonane, A.~Cintins, A.~Kalinko, R.~Chernikov, A.~Kuzmin, {Probing the
		thermochromic phase transition in CuMoO$_4$ by EXAFS spectroscopy}, Phys.
	Status Solidi B 255 (2018) 1800074.
	\newblock \href {https://doi.org/10.1002/pssb.201800074}
	{\path{doi:10.1002/pssb.201800074}}.
	
	\bibitem{Pudza2021}
	I.~Pudza, A.~Kalinko, A.~Cintins, A.~Kuzmin, {Study of the thermochromic phase
		transition in CuMo$_{1-x}$W$_x$O$_4$ solid solutions at the W L$_3$-edge by
		resonant X-ray emission spectroscopy}, Acta Mater. 205 (2021) 116581.
	\newblock \href {https://doi.org/10.1016/j.actamat.2020.116581}
	{\path{doi:10.1016/j.actamat.2020.116581}}.
	
	\bibitem{P65}
	E.~Welter, R.~Chernikov, M.~Herrmann, R.~Nemausat, {A beamline for bulk sample
		x-ray absorption spectroscopy at the high brilliance storage ring PETRA III},
	AIP Conf. Proc. 2054 (2019) 040002.
	\newblock \href {https://doi.org/10.1063/1.5084603}
	{\path{doi:10.1063/1.5084603}}.
	
	\bibitem{Ravel2005}
	B.~Ravel, M.~Newville, {ATHENA, ARTEMIS, HEPHAESTUS: data analysis for X-ray
		absorption spectroscopy using IFEFFIT}, J. Synchrotron Radiat. 12 (2005)
	537--541.
	\newblock \href {https://doi.org/10.1107/S0909049505012719}
	{\path{doi:10.1107/S0909049505012719}}.
	
	\bibitem{Kuzmin2014}
	A.~Kuzmin, J.~Chaboy, {EXAFS and XANES analysis of oxides at the nanoscale},
	IUCrJ 1 (2014) 571--589.
	\newblock \href {https://doi.org/10.1107/S2052252514021101}
	{\path{doi:10.1107/S2052252514021101}}.
	
	\bibitem{Timoshenko2014analysis}
	J.~Timoshenko, A.~Anspoks, A.~Kalinko, A.~Kuzmin, {Analysis of extended x-ray
		absorption fine structure data from copper tungstate by the reverse Monte
		Carlo method}, Phys. Scr. 89 (2014) 044006.
	\newblock \href {https://doi.org/10.1088/0031-8949/89/04/044006}
	{\path{doi:10.1088/0031-8949/89/04/044006}}.
	
	\bibitem{Timoshenko2015}
	J.~Timoshenko, A.~Anspoks, A.~Kalinko, I.~Jonane, A.~Kuzmin, {Local structure
		of multiferroic MnWO$_4$ and Mn$_{0.7}$Co$_{0.3}$WO$_4$ revealed by the
		evolutionary algorithm}, Ferroelectrics 483 (2015) 68--74.
	\newblock \href {https://doi.org/10.1080/00150193.2015.1058687}
	{\path{doi:10.1080/00150193.2015.1058687}}.
	
	\bibitem{Ankudinov1998}
	A.~Ankudinov, B.~Ravel, J.~Rehr, S.~Conradson, {Real-space multiple-scattering
		calculation and interpretation of x-ray-absorption near-edge structure},
	Phys. Rev. B 58 (1998) 7565.
	\newblock \href {https://doi.org/10.1103/PhysRevB.58.7565}
	{\path{doi:10.1103/PhysRevB.58.7565}}.
	
	\bibitem{Hedin1971}
	L.~Hedin, B.~I. Lundqvist, {Explicit local exchange-correlation potentials}, J.
	Phys. C: Solid State Phys. 4 (1971) 2064.
	\newblock \href {https://doi.org/10.1088/0022-3719/4/14/022}
	{\path{doi:10.1088/0022-3719/4/14/022}}.
	
	\bibitem{Timoshenko2009}
	J.~Timoshenko, A.~Kuzmin, {Wavelet data analysis of EXAFS spectra}, Comput.
	Phys. Commun. 180 (2009) 920--925.
	\newblock \href {https://doi.org/10.1016/j.cpc.2008.12.020}
	{\path{doi:10.1016/j.cpc.2008.12.020}}.
	
	\bibitem{Daszykowski2007robust}
	M.~Daszykowski, K.~Kaczmarek, Y.~Vander~Heyden, B.~Walczak, {Robust statistics
		in data analysis -- A review: Basic concepts}, Chemom. Intell. Lab. Syst. 85
	(2007) 203--219.
	\newblock \href {https://doi.org/10.1016/j.chemolab.2006.06.016}
	{\path{doi:10.1016/j.chemolab.2006.06.016}}.
	
	\bibitem{Sevillano1979}
	E.~Sevillano, H.~Meuth, J.~Rehr, {Extended x-ray absorption fine structure
		Debye-Waller factors. I. Monatomic crystals}, Phys. Rev. B 20 (1979) 4908.
	\newblock \href {https://doi.org/10.1103/PhysRevB.20.4908}
	{\path{doi:10.1103/PhysRevB.20.4908}}.
	
\end{thebibliography}
\end{document}